\documentclass[%
 reprint,
 amsmath,amssymb,
 aps,
twocolumn,prl]{revtex4}

\usepackage{graphicx}
\usepackage{dcolumn}
\usepackage{bm}

\begin{document}

\preprint{APS/123-QED}

\title{Statistical early-warning indicators  based on  Auto-Regressive Moving-Average  processes }

\author{Davide Faranda, B\'ereng\`ere Dubrulle}
 \email{davide.faranda@cea.fr}
\affiliation{%
Laboratoire SPHYNX, Service de Physique de l'Etat Condens\'e, DSM,
CEA Saclay, CNRS URA 2464, 91191 Gif-sur-Yvette, France
}
 
\author{Flavio Maria Emanuele Pons}
\affiliation{%
Department of Statistics, University of Bologna, 
Via delle Belle Arti 41, 40126 Bologna, Italy .}


\begin{abstract}
We address the problem of defining early warning indicators of critical transition. To this purpose, we fit the relevant time series through a class of linear models, known as  Auto-Regressive Moving-Average (ARMA($p,q$)) models. We define two indicators representing the total order and the total persistence of the process, linked, respectively, to the shape and to the characteristic decay time of the autocorrelation function of the process. We successfully test the method to detect transitions in a Langevin model and a 2D Ising model with nearest-neighbour interaction. We then apply the method to complex systems, namely for dynamo thresholds and financial crisis detection.
\end{abstract}

\pacs{Valid PACS appear here}
\maketitle


 Many experimental or natural systems undergo  {\it critical transitions\/} -   sudden shifts from one to another dynamical regime. In some instances,  {\it e.g.} global changes in climate science, species extinction in ecology, spin glasses, it is of crucial importance to build early warning indicators, i.e. estimates of the transition threshold based on finite time-series corresponding to situations where the bifurcation did not happen yet. The statistical approach to this issue traditionally involves so-called indicators of criticality~\cite{scheffer2009early,dakos2012methods}. They are based on specific properties of ideal statistical systems (such as the Langevin or Ising model) near the transition: critical slowing down, modifications of the auto-correlation function or of the fluctuations~\cite{farandamanneville}, increase of variance and skewness~\cite{kuehn2011mathematical}, diverging susceptibility~\cite{monchaux2009karman,berhanu2009bistability,miralles}, diverging correlation length. However, it is known that, in some complex systems, these indicators fail to detect the transition: in spin glasses, no diverging correlation length has been found so far, and one has to resort to finer statistical tools (such as four point dynamical susceptibility~\cite{franz1999dynamical,parisi1997approach}) to detect transitions.
 In addition, traditional early warning  indicators  may be  inapplicable in datasets containing a small number of observations (see {\it e.g.} \cite{faranda2011numerical,faranda2013recurrence}), which is usually the case in several applications where the experiment  is unique (as in financial or climate time series),  difficult to repeat or to sample for a long time (as for atmospheric measurements, laboratory turbulence, etc). 
 This suggests that indicators based on single statistical properties may not be sufficient for detecting transitions in complex systems, so that one should rather consider all the information contained in the finite-time series.\  
 
 The main idea of the present letter is therefore to introduce a class of indicators of critical transitions based on a statistical model for the observed data when approaching a tipping point. To be interesting for applications, these indicators must satisfy certain properties:  i) they must generalize the well-know indicators based on single statistical properties and ii) they must be built using a statistical model that is simple to implement and works for limited data set. Here, we show that these properties are satisfied for indicators based on the auto-regressive moving-average processes of order $p,q$ ARMA($p,q$), modeling  a time series $X_t(\lambda)$, experiencing a transition at $\lambda=\lambda_c$. In the first part of the paper, we first recall some basics on ARMA($p,q$) modeling and define corresponding early-warning indicators. We then check that these indicators are able to detect the transition in simple theoretical models, such as Langevin double-well model or Ising model. We then apply our indicators to the analysis of complex systems for the detection of turbulent dynamo thresholds and financial crisis.\\

\paragraph*{Theory.}    
 Let us consider a series of observations $X_t$ of an observable with unknown underlying dynamics, controlled by a parameter $\lambda$. We further assume that for $\lambda<\lambda_c$ the time series $X_t(\lambda)$ represents a stationary phenomenon. The critical threshold $\lambda_c$ is defined through the condition that for  $\lambda\geq\lambda_c$, there is a bifurcation, in the sense that  there exists no smooth transformation of the physical measure through the transition.  Since  $X_t(\lambda)$ is stationary, we may then model it by an ARMA$(p(\lambda),q(\lambda))$ process such that   for all $t$:
\begin{equation}
X_t(\lambda) = \sum_{i=1}^p \phi_i(\lambda) X_{t-i} + \varepsilon_{t} + \sum_{j=1}^q \theta_j(\lambda) \varepsilon_{t-j}
\label{ARMA1}
\end{equation}
with $\varepsilon_t  \sim WN(0, \sigma^2)$ - where $WN$ stands for white noise - and the polynomials $\phi(z) = 1 - \phi_1 z_{t-1} - \cdots - \phi_p z_{t-p}$ and $\theta(z) = 1 - \theta_1 z_{t-1} - \cdots - \theta_q z_{t-q}$, with $z \in \mathbb{C}$, have no common factors. Notice that, hereinafter, the noise term $\varepsilon_t$ will be assumed to be a white noise, which is a very general condition \cite{brockwell2009time}. For a general stationary time series, this model is not unique.  However there are several standard procedures for selecting the model which fits at best the data. The one we exploit in this paper is the Box-Jenkis procedure \cite{box_etal-springer-1970}. We chose the lowest $p$ and $q$ such that the residuals of the series filtered by the process ARMA($p,q$) are delta correlated. This  fixes $p$ and $q$, and thus our statistical model. There are other model selection procedures based on information criteria (Bayesan or Akaike information criteria).  Unfortunately, in our case none of them gives clear indications for discriminating the model to be used as it is not clear which range $p$ and $q$ must be tested to get reliable results.  Intuitively, $p$ and $q$ are related to memory lag of the process, while  the  coefficients $\phi_i(\lambda)$ and $\theta_i(\lambda)$ represent the persistence: the higher their sum (in absolute value), the slower the system is forgetting its past history. In the sequel, we now present early warning indicators based on these parameters.

\paragraph{Early-warning indicators.} 
Far from the transition, the time series of a generic physical observable can be described by an ARMA($p,q$) model with a reasonably low number of $p,q$ parameters and coefficients. On the other hand, for $\lambda \to \lambda_c$, the critical value corresponding to a transition, the statistical properties (such as the shape and/or the persistence of the autocorrelation function) of the system change, leading to different characteristics of the ARMA($p,q$) model which can describe the data series or to an inadequacy of the model itself.  
Specifically, several changes in the dynamics occur near the transition, as the system is allowed to explore a larger portion of the phase space with higher excursions in the direction of the new stable state. First, the distributions of the observables become skewed towards the maxima or the minima, depending on the direction of the shift. Second, the system may experience a \textit{critical slowing down} with diverging memory effects.  This phenomenon is traditionally quantified by the autocorrelation function (ACF) of the time series $X_t$ defined (assuming a zero-mean process) as:  
\begin{equation}\label{acf}
\mbox{ACF}(h) = E[X_{t+h} X_t]/E[X_t^2].
\end{equation}
Far from the transition, the ACF tends to be 0 after a finite number of lags $\bar{h}$. As $\lambda \to \lambda_c$, critical increase of memory of the system makes $\bar{h} \to \infty$. The ARMA($p(\lambda),q((\lambda)$) model of the corresponding time series will then be characterized by two properties:
\begin{itemize}
\item  $\sum_{i=1}^p |\phi_i|$ and $\sum_{j=1}^1 |\theta_j|$ increase  for $\lambda \to \lambda_c$    as the direct consequence of $\bar{h} \to \infty$.
\item  $p+q$ increases for  $\lambda \to \lambda_c$ because of additional new time scales associated   to the  trajectories moving towards  the potential barrier between the two attracting states.
\end{itemize}
This rather simple observation allows us to define two indicators:  $\mathcal{O(\lambda)}=p(\lambda)+q(\lambda)$, which  diverges for $\lambda\to \lambda_c$, and  the  total persistence of the process: $$ \mathcal{R(\lambda)}= \sum_{i=1}^p|\phi_i(\lambda)| +\sum_{i=1}^q|\theta_i(\lambda)|$$  which also show a divergent behavior at the transition, unless $\mathcal{O(\lambda)}=1$. In this latter case  $\mathcal{R(\lambda)}=\phi_1 \to 1$ for $\lambda \to \lambda_c$.  
These indicators present several advantages with respect to  the traditional ones reported, for example, in \cite{scheffer2009early}. First,  by computing the functional form for $p(\lambda)$ and $q(\lambda)$ and for the coefficients $\phi_i(\lambda)$ and $\theta_i(\lambda)$  one has also an effective   statistical toy model for describing the phenomenon and to produce data with analogous statistical properties. This may  be very useful for series or data which can hardly be reproduced (laboratory experiments) or integrated  by new observations (climate datasets, stock market titles).  Second, if several series at different $\lambda$'s are available,  one can extrapolate the characteristics of the process at  not yet measured $\lambda$'s. This   property can be very useful for devising new experiments  knowing  the possible location of the transitions.   Third, if the transition is marked   by the fact that $\mathcal{R}(\lambda) \to \mathcal{O(\lambda)}$  rather than by  a divergence of $\mathcal{O(\lambda)}$, one  may argue that the potential landscape for the observable $X$ does not change significantly when approaching the transition and therefore a Langevin reduction to a double well system is possible. If, on the contrary, the order changes significantly approaching the transition, such a low dimensional reduction is not appropriate and one should be very careful in pursuing such a model as shown, for a relevant climatic example, in \cite{lucarini2012bistable}.\\

\paragraph*{A toy model for critical transitions.}

\begin{figure}
\includegraphics[width=0.5\textwidth]{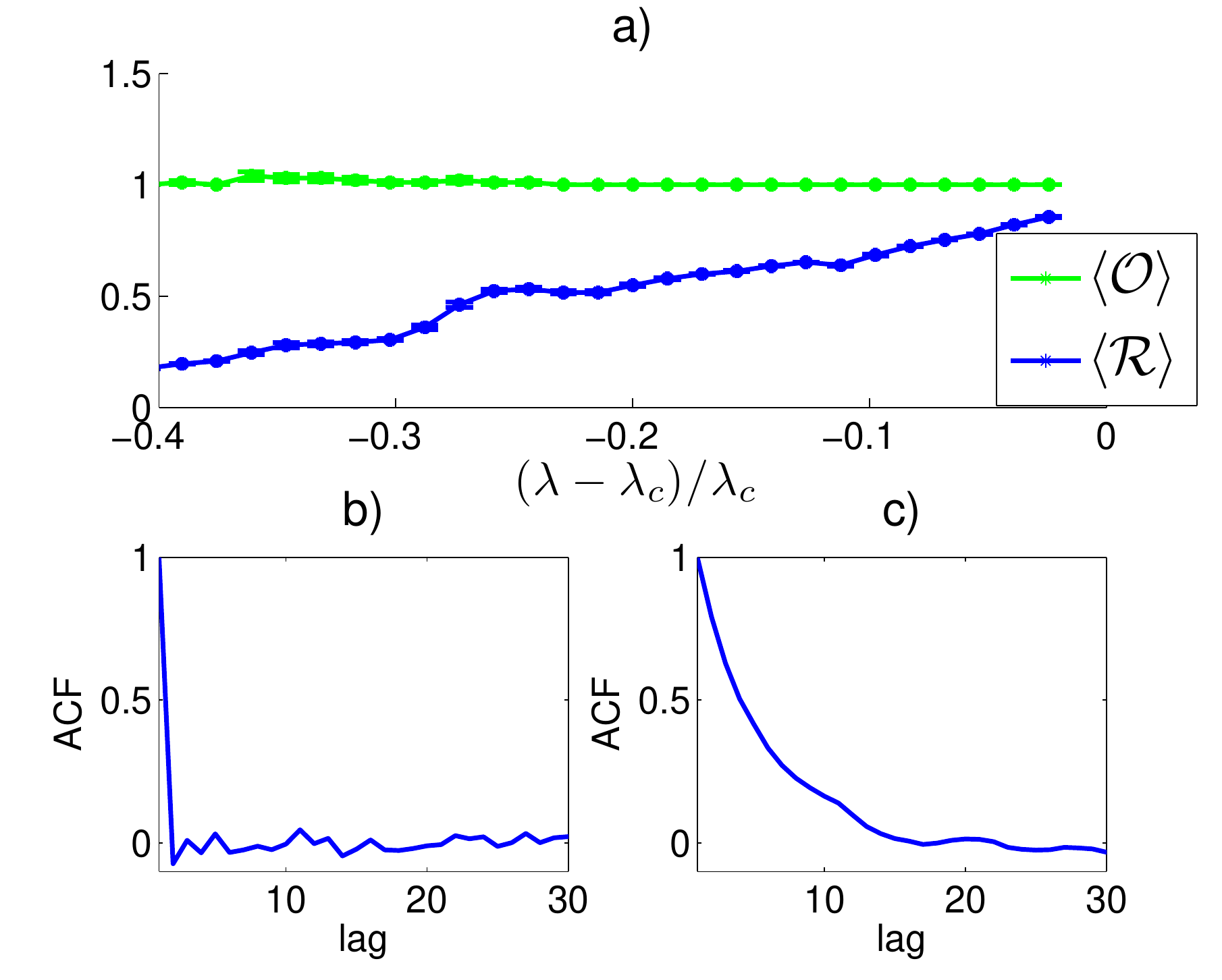}
\caption{  (a):$\langle\mathcal{O}\rangle$ and $\langle\mathcal{R}\rangle$ for the system defined in Eq.~\ref{dw}. Each error-bar represents the average of 15 realizations and the standard deviation of the mean.  b)   ACF for $(\lambda-\lambda_c)/\lambda_c=-0.271$. c) ACF for $(\lambda-\lambda_c)/\lambda_c=-0.05$ .} 
\label{lan}
\end{figure}

 We start considering a classical system featuring bistability under the effect of random noise, i.e.
\begin{equation}
{\rm d}X=-V'(X){\rm d}t +\epsilon {\rm d}W
\label{dw}
\end{equation}
with potential  $V(X)=aX^4 -bX^2+\lambda X$, where  $\lambda\geq0$  and  $W$ is a Wiener process with unit variance. 
We consider system (\ref{dw}) for values of $\lambda$ such that, in the deterministic limit, it features two stable fixed points ($\bar X_1<0$ and $\bar X_2>0$) and an unstable fixed point $\tilde X$.
The asymptotic behavior of the system can be assessed in terms of the solution of a Fokker-Plank equation~\cite{risken1989fokker}.
Here we are rather interested  in the finite-time behavior and we consider only the simulations such that the noise does not push the system across the bifurcation, {\it i.e.} the system is confined in one of the two wells. For this system we perform the following numerical experiments: for each value of $\lambda$ we compute an ensemble of 500 trajectories $X(\lambda)$  finding, for each of them, the best ARMA($p(\lambda),q(\lambda)$) in the sense specified by the Box-Jenkins procedure \cite{box_etal-springer-1970}. Then,  $\langle\mathcal{O}(\lambda)\rangle$ and $\langle\mathcal{R}(\lambda)\rangle$   have been computed,  here $\langle\cdot\rangle$ stands for the ensemble average.\\
 In Fig. \ref{lan}-a we report the results of this analysis, which clearly show that the average order is not affected in this case and $\langle\mathcal{O}(\lambda)\rangle\simeq 1$, whereas the transition is well highlighted  by $\langle\mathcal{R}(\lambda)\rangle$ which approaches $\langle\mathcal{O}(\lambda)\rangle=1$ for $\lambda \to \lambda_c$ . There is a simple way to understand this behavior by linking the  orders $p,q$  and of the coefficients $\phi_i$ to the  autocorrelation function ACF (see \cite{brockwell2009time}- Chapter 3 for more details).  The orders are linked to the functional form of the ACF whereas the values of $\phi_i$  depend on the decay rate. In the case of  system given by Eq.~\ref{dw}, the  shape of the ACF is exponential both far from the transition (Fig \ref{lan}-b)  and when  approaching it (Fig \ref{lan}-c). However, in the latter case, the decay is much slower, this causing the increase  $\mathcal{R}(\lambda) \to 1$.

\begin{figure}
\includegraphics[width=0.5\textwidth]{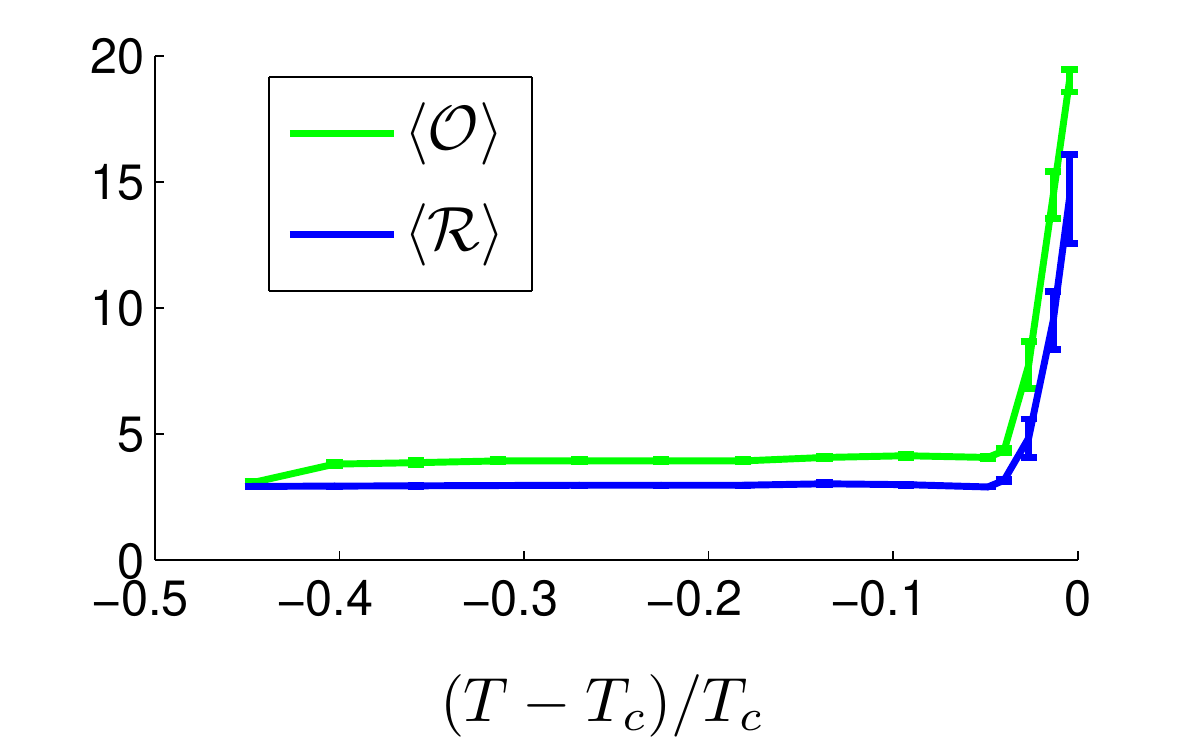}
\caption{  $\langle\mathcal{O}\rangle$  and  $\langle\mathcal{R}\rangle$ for the system defined in Eq.~\ref{ising}, L=256. Each error-bar represents the average of 15 realizations and the standard deviation of the mean. } 
\label{isingfig}
\end{figure}

\paragraph*{The Ising model.} 
As a second test of the indicator, we consider a classical 
 2-D Ising dynamics with a nearest-neighbor interaction on a square lattice  of size $L$. At each site $j$, a discrete spin is allowed to have two values $\sigma_j \in \{ +1,-1 \}$. The energy of the configuration is given by the Hamiltonian: 
\begin{equation}
\mathcal{H}=-J\sum_{\rm neighbors}\sigma_i \sigma_j
\label{ising}
\end{equation}
under the interaction $J$. We consider only the case $J=1$ and  evolve the system by using Metropolis algorithm \cite{haario2001adaptive}. A second order  phase transition is expected at the temperature $T=T_c=2/\ln(1+\sqrt{2}) \simeq 2.269$.  To apply our early warning indicators, we performed 100 simulations for a square lattice  of size $L=256$   at different $T>T_c$. We checked that our results  do not depend sensitively on the size of the lattice, provided that $L>128$.  After discarding 100000 time iterations  necessary to reach a  clearly identifiable stationary state, for each temperature, an ensemble of 15 time series consisting of 200000 time units of $\mathcal{H}(t,T)$  is analyzed with the procedure described above. Stationarity has been tested performing a Dickey-Fuller test on each time series. The results for  $\langle\mathcal{O}\rangle$ and $\langle\mathcal{R}\rangle$ are  reported in Fig. \ref{isingfig}. It is evident that   $\langle\mathcal{O}\rangle$ and $\langle\mathcal{R}\rangle$     increase when moving towards the critical temperature $T_c$.   In this case, not only the persistence of the correlations  $\mathcal{R}$, but also the number of terms  $\mathcal{O}$ necessary to describe the process increases. This means that the transition cannot be modeled by a simple Langevin equation as other time scales become important. In other words, this transition is associated to a non-trivial unknown potential landscape.

\paragraph*{An example of complex system.}
 
\begin{figure}
\includegraphics[width=0.45\textwidth]{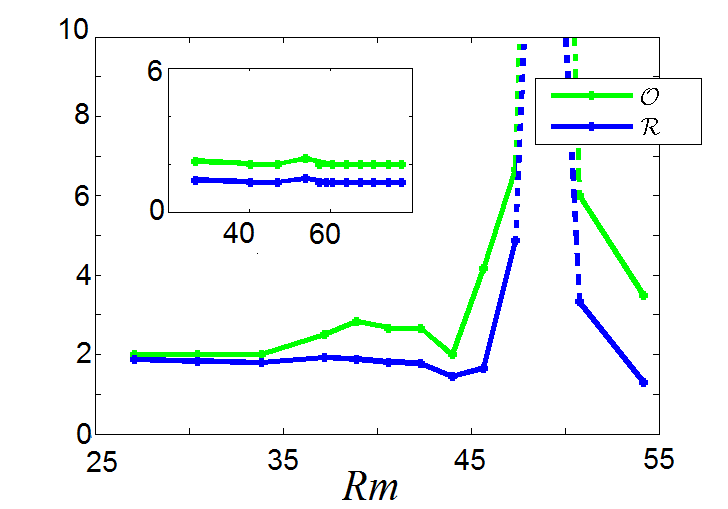}
\caption{ Averaged $\mathcal{O}$ and $\mathcal{R}$ for the Von Karman - Sodium experiment. Solid lines refer to the experiments for increasing values of $Rm$. Inset: same as the main figure but for a configuration where the dynamo has not been observed.}
\label{VKS}
\end{figure}

\begin{figure}
\includegraphics[width=0.5\textwidth]{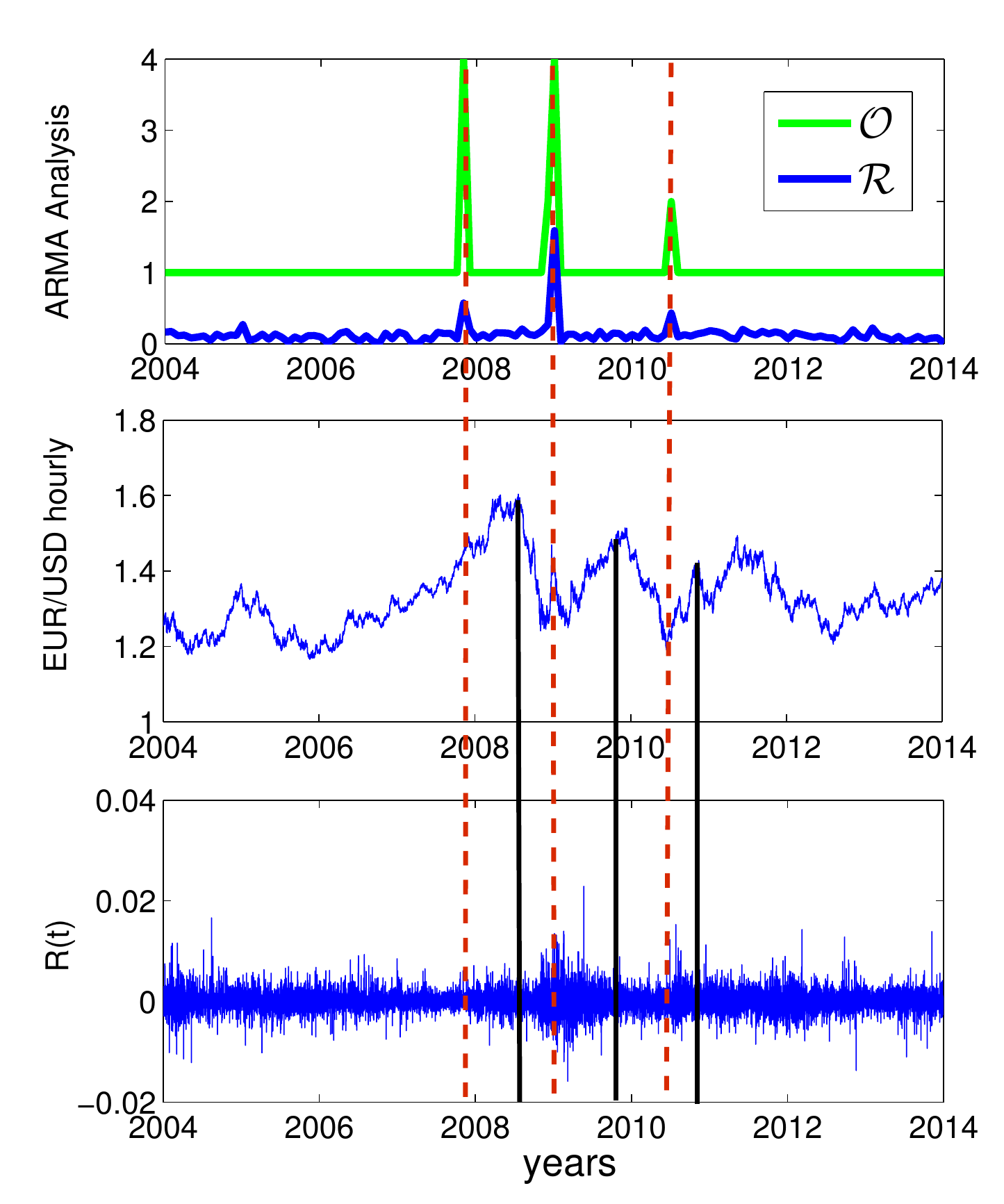}
\caption{ Upper panel: Average $\mathcal{O}$ and $\mathcal{R}$ for the log-return series $R(t)$ of the EUR/USD hourly exchange rate. Central panel:  EUR/USD hourly exchange rate. Lower panel: $R(t)$. Red  dotted lines correspond to early warning of the crisis. Black continuous lines correspond to actions taken by the ECB.}
\label{eur}
\end{figure}

Up to now we have analyzed toy systems, extensively studied both analytically and numerically and for which the threshold are analitically predictable. However, interesting systems, such as turbulence or finance, lie on another level of complexity and one naturally wonders whether the technique provides reliable results.  We focus on  the data of the  Von K\'arm\'an Sodium (VKS) experiment, a successful attempt  to get  a transition to dynamo in a laboratory turbulent liquid-metal experiment.  The control parameter for the transition is the   magnetic Reynolds number $Rm$. The interesting characteristic of this experiment is that several dynamo and no dynamo configurations  have been  obtained by changing the material of the impellers and of the  cylinder \cite{miralles,boisson2012symmetry}.  
Here we focus on two different configurations: (i) one producing  a well-documented stationary dynamo at $Rm\approx 44$, thereby providing a fair test of our method and (ii) one that failed to produce the dynamo within the accessible values of $Rm$. The time series is constructed using  the modulus of the magnetic field $\vert{B}\vert(t)$ as a function of time $t$, measured within six detectors in the core of the vessel. From this, we extract  the quantities $\mathcal{O}$ and $\mathcal{R}$, averaged over the six detectors. The results are plotted  in Fig. \ref{VKS}: the main figure for the configuration (i), the inset for the configuration (ii). They depend quantitatively on the sensors chosen, but not qualitatively as the transition is always detected at the same $Rm$.  The transition is very net  in terms of divergence of  $\mathcal{O}$ and $\mathcal{R}$ and can be located at $Rm_f=47$, when the dynamo is observed. In the non-dynamo case, no sign of transition is visible. \\ 
\paragraph*{Financial crisis.}
We conclude by discussing the performances of the ARMA early warning indicators applied to the EUR/USD exchange rate hourly datasets (Figure \ref{eur}-b). The chosen observable is the log-return of the time series, a quantity commonly examined in finance as the series obtained this way do not contain non-stationarities:
$$ R(t)=log(X_t)-log(X_{t-1}).$$
Here $X_t$ is the EUR/USD exchange hourly rate. Monthly values for $\mathcal{O}$ and $\mathcal{R}$ have been extracted  from the time series of $R(t)$ and results are shown in figure \ref{eur}-a. Our analysis can be safely performed on these series since they are stationary, as it results from  the Dickey-Fuller test \cite{dickey1981likelihood}. The technique  clearly points to three distinct warnings (marked by the red dotted lines). Interestingly, they are followed  after a  few months  delay, by official warnings of the European Central Bank (ECB). The first warning corresponds to the Sub-prime American crisis, the second to the Greek crisis, and the third one to the Irish crisis. The crisis for the real market falls immediately after the ECB announcements. If we compare these results with the ones arising from physical systems, the warning seems to appear \textit{too early}.  We may argue that   indicators which provide similar warnings are available also at the ECB. The time between the early warning discover and the ECB announcements may serve to the ECB  for trying corrections and avoid an immediate financial crisis which is   announced only when the crisis itself is unavoidable.  Similar behaviors have been discovered for early warning indicators applied to financial datasets, as reported in \cite{fischer2003financial,davis2008comparing} . 
\paragraph*{Discussion.}
In this work we have proposed a new  method to detect early warnings of critical transitions via a statistical approach which allows also for incorporating the information of several statistical indicators analysed in \cite{dakos2012methods}. Here, we   exploit the properties of linear ({\itshape i.e.} stationary and invertible) autoregressive moving-average processes, denoted ARMA($p(\lambda), q(\lambda))$, being $\lambda$ the system control parameter. 
More specifically, we have defined two indicators representing the total order and the total persistence of the stochastic process. An increase of the former is indicative of the impossibility to represent the data series in a parsimonious way, thus leading to the idea that the linearity hypothesis fails and the decay of the autocorrelation function of the process is no longer exponential; an increase of the latter is linked to a longer decay time of the correlation, and can be due to the increase of the total order or just of the persistence itself. The two phenomena are very different and, up to our knowledge, the traditional indicators exploit only the increase of the memory of the system (critical slowing down) to identify the threshold $\lambda_c$. Here, the possibility of detecting substantial modifications to the shape of the autocorrelation function via the change in the orders $p,q$, allows for understanding whether reductions to simple low dimensional models are relevant or not for describing the dynamics. We have combined these two indicators to detect critical transitions both in models and in real systems. In all the cases considered, the behavior of the indicators has shown to be an effective way to investigate the proximity of the system to a critical transition; thus, they seem to be a useful tool to study critical transitions, since their estimation involves well-known, standard statistical techniques characterized by a low computational cost and applicable to relatively short time series.\\
The application to finance seems promising. It would be interesting to extend this approach to other financial indicators 	as well as to climate data. On a theoretical level, one could use the technique to understand how transitions are modified when systems originally in equilibrium are driven out of equilibrium by forcing-dissipation mechanisms, starting from conceptual toy model of  out-of-equilibrium  Ising dynamics \cite{pleimling2010convection,faranda2013non}.

\bibliography{armatrans}

\end{document}